\documentclass[namedreferences,fleqn]{spackap}
\usepackage{klups}
\usepackage{psfig}
\begin{document}
\bibliographystyle{klunamed}
\begin{article}
\newcommand\apj[6] { #1: #2, `#3', {\it Astrophys.~J.} {\bf#4}, #5--#6.}
\newcommand\mnras[6] { #1: #2, `#3', {\it Monthly Notices Royal Astron. Soc.} 
{\bf#4}, #5--#6.}
\newcommand\aap[6] { #1: #2, `#3', {\it Astron. Astrophys.} {\bf#4}, #5--#6.}
\newcommand\icarus[6] { #1: #2, `#3', {\it Icarus} {\bf#4}, #5--#6.}
\newcommand\wig[1]{\mathrel{\hbox{\hbox to 0pt{%
\lower.5ex\hbox{$\sim$}\hss}\raise.4ex\hbox{$#1$}}}}
\begin{opening}
\title{ATOMIC DEUTERIUM/HYDROGEN IN THE GALAXY\vspace{-0.25cm}}
\author{Jeffrey L. \surname{LINSKY}\vspace{0.1cm}}
\institute{JILA, University of Colorado and NIST, Boulder, CO 80309-0440 
USA\vspace{-0.2cm}}
\begin{ao}
Jeffrey L. Linsky, JILA, University of Colorado, Boulder, CO 80309-0440, 
USA; \ {\tt jlinsky@jila.colorado.edu}
\end{ao}
\runningauthor{Jeffrey L. Linsky} 
\runningtitle{ATOMIC D/H IN THE GALAXY}
\date{Received: ; Accepted in final form: } 
\begin{abstract}
An accurate value of the deuterium/hydrogen (D/H) 
ratio in the local interstellar medium (LISM) and
a better understanding of the D/H variations with position in the Galactic disk
can provide essential information on the primordial D/H ratio in the Galaxy at the
time of the protosolar nebula, and the amount of astration and mixing in the
Galaxy over time. Recent measurements have been obtained with UV spectrographs
on FUSE, HST, and IMAPS using hot white dwarfs, OB stars, and late-type stars
as background light sources against which to measure absorption by D and H in
the interstellar medium along the lines of sight. Recent analyses of FUSE
observations of seven white dwarfs and subdwarfs provide a weighted mean value
of D/H = $(1.52 \pm 0.08) \times 10^{-5}$ ({\bf $15.2\pm 0.8$} ppm), consistent
with the value of $(1.50 \pm 0.10) \times 10^{-5}$ ({\bf $15.0\pm 1.0$} ppm)
obtained from analysis of lines of sight toward nearby late-type stars. Both
numbers refer to the ISM within about 100 pc of the Sun, which samples warm
clouds located within the Local Bubble. Outside of the Local Bubble at
distances of 200 to 500 pc, analyses of far-UV spectra obtained with IMAPS
indicate a much wider range of D/H ratios between 0.8 to 2.2 ppm, providing
information on inhomogeneous astration in the Galactic disk.
\end{abstract}
\end{opening}
\setcounter{page}{1} 
\volume{95/96}

\section{Why are Accurate Measurements of the D/H Ratio Important?}

Measurements of D/H, the number ratio of deuterium in all forms to hydrogen in
all forms, are important for at least two reasons. First, an accurate
measurement of the primordial ratio, (D/H)$_{prim}$, counts the number of
baryons in the universe to determine the ratio $\Omega_B$ of the baryon density
to the closure density, and tests our assumptions concerning nucleosynthesis
during the first 100--1,000 seconds of the universe (e.g.,
\opencite{Burlesetal01}). Deuterium is our best probe of primordial
nucleosynthesis because theory predicts that D was formed only in the very
early universe, D is the easiest isotope to be destroyed by nuclear reactions
in stars (astration), and $\Omega_B$ is a very sensitive single-valued function
of (D/H)$_{prim}$. The best approximation to (D/H)$_{prim}$ would be an
accurate measurement of D/H in gas where there has been little chemical
fractionation or star formation, as indicated by very low metal abundances, but
such measurements remain difficult. 

Second, measurements of D/H in different locations in our Galaxy will provide
an accurate test of the assumptions underlying Galactic chemical evolution
models. A major problem in astrophysics is understanding how galaxies evolve
and, in particular, how the chemical element abundances evolve. In broad
overview, we know that stars form out of gas clouds and over time they destroy
D, create metals, and return some of this deuterium-poor and metal-rich
material to the ISM by winds and supernova explosions. The detailed rates for
these processes depend on the initial stellar masses. Thus with time D/H should
decrease and metal abundances should increase. Theoretical models for Galactic
chemical evolution rest on many assumptions that measurements of D/H in
different environments can test. In particular, the temporal and spatial scales
for mixing in the ISM are poorly known and likely depend on the magnetic field,
which is also poorly known.

\section{What is the Best Way of Measuring D/H?}

While I believe that the most accurate D/H measurements are obtained from
interstellar H and D Lyman line absorption in warm interstellar gas, I first
summarize the various techniques that have been used to measure D/H in the
Galaxy: 

\begin{description}

\item[Deuterated molecules in cold interstellar clouds:] HDO/H$_2$O $\geq$ {\bf
1000} ppm\footnote{I express D/H ratios in parts per million (ppm) in {\bf bold 
face} to facilitate intercomparisons.}
and other deuterated molecules also show very high abundances. Since
deuterated molecules are more tightly bound than nondeuterated molecules, the
small difference in the binding energies divided by $kT$ can be large at cold
temperatures (10--20 K). For example, the reaction HD + H$_2$O
$\leftrightarrow$ HDO + H$_2$ at low temperatures leads to HDO/H$_2$O $\gg$
D/H. Carbon molecule chemistry also creates huge overabundances of the
deuterated molecules. 

\item[HD/H$_2$ in the ISM:] In cold clouds nearly all D is tied up in HD
molecules, so HD/H$_2$ measures D/H. Measurement of the HD J = 1 $\rightarrow$
0 pure rotation line (112 $\mu$m) in the Orion Bar \cite{Wrightetal99} by the
ISO spacecraft gives D/H = {\bf 10} $\pm$ {\bf 3} ppm. This value for D/H may
not be representative of the gas, however, since HD is not self-shielded like
H$_2$ and will have a higher photodissociation rate from stellar and diffuse
UV radiation. 

\item[Balmer-$\alpha$ line in the Orion Nebula:]
\inlinecite{Hebrardetal00} first detected narrow deuterium Balmer-$\alpha$ and
Balmer-$\beta$ emission lines. Accurate measurements of the D/H ratio from the
Balmer lines is difficult, however, because the D Balmer lines are fluorescent
lines pumped by the hot star continuum, whereas the H Balmer lines are
recombination lines (cf. \opencite{Odelletal01}). 

\item[Hyperfine structure line:] The most recent search for the 92 cm (327 MHz)
deuterium line in the ISM toward the Galactic anticenter yields a possible
detection \cite{Chengaluretal97} with D/H = ${\bf 39\pm 10}$ ppm. 

\item[D/H in the Sun:] A search for D Balmer-$\alpha$ emission at --1.785~\AA\
relative to H Balmer-$\alpha$ \cite{Beckers75} gives an upper limit of D/H $<$
{\bf 0.25} ppm. This very low value for D/H is consistent with the burning of D
deep in the convective zone and the mixing of this D-depleted gas throughout
the solar atmosphere. 

\item[D/H in the solar system:] In his review paper, \inlinecite{Robertetal00}
list for D/H in comets {\bf 300} ppm, meteorites {\bf 80--1000} ppm, Jupiter
and Saturn {\bf 25} ppm, Uranus and Neptune {\bf 60} ppm. The standard
explanation is that the initially highly deuterated water and other molecules
become less deuterated with time by isotopic exchange with H$_2$ at warmer
temperatures. Terrestrial water also started with a very high D/H ratio and
subsequently reached its present ratio of HDO/H$_2$O = {\bf 150} ppm via
partial isotopic re-equilibrium with warm H$_2$. 

\end{description}
\vspace{-11pt}

\section{Measuring D/H with UV Spectra from HST}

The Goddard High Resolution Spectrograph (GHRS) and the Space Telescope Imaging
Spectrograph (STIS) instruments on HST with resolution of $\leq 3$ km~s$^{-1}$
are providing beautiful spectra of interstellar Lyman-$\alpha$ absorption 
with which to measure the column densities N(D~I) and N(H~I) and thus D/H.
Several serendipidous results have already emerged from this analysis.\\ 


\noindent \underline{Virtues of this approach:}

\begin{itemize}
\item[$\bullet$] Since no molecules are present in the warm ($T\approx
7,000$~K) ISM clouds, there is no chemical fractionation and the fractional
ionization of H and D are the same. Thus N(D~I)/N(H~I) equals the D/H ratio in
these warm clouds. 

\item[$\bullet$]  For lines of sight through the LISM, N(H~I) $\sim 10^{18}$ --
$10^{20}$ cm$^{-2}$ and N(D~I) $\sim 10^{13}$ -- $10^{15}$ cm$^{-2}$. Thus for
either Lyman-$\alpha$ or higher Lyman series lines, the D line has measurable
opacity while the corresponding H line is not too optically thick to absorb
completely the D line located at --82 km s$^{-1}$. The ``horizon'' set by the
H~I column density at which the saturated core of the interstellar H absorption
is as wide as 82 km s$^{-1}$ is $6\times 10^{18}$ cm$^{-2}$ for Lyman-$\alpha$,
$4\times 10^{19}$ cm$^{-2}$ for Lyman-$\beta$, and larger for the higher Lyman
lines. 

\end{itemize}

\noindent \underline{Problems with this approach:}

\begin{itemize}
\item[$\bullet$]  For many lines of sight, overlapping velocity components may
permit one to measure (D/H)$_{total}$, but not D/H for each component
separately. 

\item[$\bullet$]  Low column density cloudlets of hydrogen that are Doppler
shifted with respect to the main interstellar absorption feature add to the
saturated H Lyman line absoption but have insufficient opacity to be detected
in lines of D or any metal. When not included in the analysis, this
``invisible'' hydrogen can lead to large errors in N(H~I) and
thus the D/H ratio \cite{Lemoineetal02}. 

\end{itemize}
\vspace{11pt}

As an example of the complexities in the data analysis and the serendipidous
results that have emerged from measuring the D/H ratio for the lines of sight
to the nearby stars, I summarize recent studies (cf. \opencite{LinskyWood96}
and \opencite{Woodetal01}) of the short (1.3~pc) lines of sight to the triplet
$\alpha$~Centauri system: A (a G2~V star like the Sun), B (a K2 dwarf), and C
(Proxima Centauri, an M dwarf). 

\begin{itemize}

\item[$\bullet$] The interstellar Fe~II and Mg~II resonance lines formed in the
lines of sight to these stars show absorption at only one velocity, indicating
that there is only one warm cloud, the so-called G (or Galactic Center) Cloud
along this simple line of sight. However, the central velocity of the H
Lyman-$\alpha$ absorption is redshifted by 2.2 km s$^{-1}$ relative to the D
Lyman~$\alpha$ and metal line absorption, indicating the presence of a second
red-shifted absorber in the H line. 

\item[$\bullet$] Additional absorption on the red side of the H Lyman-$\alpha$
absorption profile (see Fig. 1) is due to the ``hydrogen wall'' in the
heliosphere produced by the interaction and charge exchange of inflowing LISM
neutral H with outflowing solar wind protons near the heliopause (e.g.,
\opencite{Zanketal01}). N(H~I)$_{H wall} \sim 0.0004\times$ N(H~I)$_{G cloud}$,
which is sufficient to explain the additional H absorption but insufficient to
provide measurable D or metal line absorption. If the H wall absorption is not
included in the analysis, then the inferred N(H~I) would be a factor of 2 too
large and the inferred D/H ratio a factor of 2 too small. 

\item[$\bullet$] Additional absorption on the blue side of the H Lyman-$\alpha$
absorption profile of $\alpha$~Cen A and B (see Fig. 1) is due to hydrogen wall
absorption in their astrospheres produced by the interaction of LISM neutral H
with their ionized stellar winds. The blue shift relative to the interstellar
absorption results from viewing the decelerated H wall from the outside. The
near absence of H wall absorption in the astrosphere of Proxima Centauri
indicates a very low mass loss rate for this star. Studies of astrospheric
absorption toward a number of nearby stars allowed \inlinecite{Woodetal02a} to
infer stellar mass loss rates as small as $10^{-15} M_{\odot} yr^{-1}$, and to
estimate the mass loss rate of the young Sun, which is important for
understanding the evolution of the Martian atmosphere. 

\item[$\bullet$] \inlinecite{Lemoineetal02} and \inlinecite{Vidalferlet02} have
argued that systematic errors in deriving N(H~I) from saturated Lyman line
absorption are much larger than previously assumed, leading to very uncertain
D/H values. Large systematic errors can indeed be present, but in several well
studied examples independent measurements of N(H~I) inferred from the shape of
the Lyman continuum absorption are in excellent agreement with the
Lyman-$\alpha$ absorption results. Since the two diagnostic techniques are very
different and the Lyman-$\alpha$ and Lyman continuum optical depths differ by a
factor of $10^6$, the agreement in N(H~I) to better than 10\% using the two
techniques indicates that the systematic errors for these lines of sight are
not large. \inlinecite{Linskyetal00} summarized the close agreement between the
two different techniques for the lines of sight to the white dwarfs HZ~43 and
G191-B2B, and for groups of late-type and white dwarf stars located within a
few degrees of each other with lines of sight through the same clouds. Examples
include the HZ~43, 31~Com, and GD~153 group, and the Capella and G191-B2B pair.

\item[$\bullet$] Analysis of Lyman-$\alpha$ absorption for 12 sightlines
through the LIC yield a mean value of D/H = {\bf 15.0$\pm$ 1.0} ppm
\cite{Linsky98} and no trend with distance to the target star (up to 100 pc) or
Galactic longitude. Other investigators have also analyzed GHRS and STIS data
using different approaches. For example, \inlinecite{Vidalmadjaretal98}
confirmed that the D/H ratio for the Capella line of sight through the LIC is
consistent with the mean LIC value. The G191-B2B line of sight has generated
more controvery, although \inlinecite{Vidalmadjaretal98},
\inlinecite{Lemoineetal02}, and \inlinecite{Sahuetal99} agree that D/H in the
LIC component is consistent with the mean value. They disagree, however, on the
value of D/H in the one or two other velocity components along the line of
sight to this star located only $69^{+19}_{-12}$~pc away. 

\end{itemize}

\begin{figure}
\centerline{\psfig{file=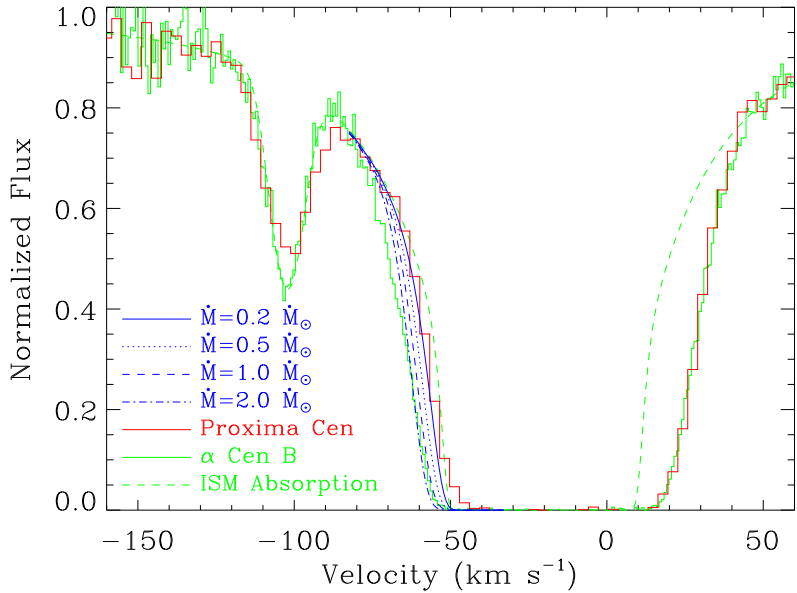,width=0.95\textwidth,clip=,angle=90}
\vspace{0.1cm}}
\caption{Comparison of the observed Lyman-$\alpha$ profiles toward
$\alpha$~Cen~B and Proxima Centauri. The dashed line is the interstellar
absorption predicted from the observed D and metal lines. The extra absorption
on the red side of the interstellar absorption (same for both stars) is due to
the H wall in the heliosphere. The extra absorption on the blue side (different
for the two stars and a function of the mass loss rate) is due to the H wall in
the astrospheres. From Wood {\it et al.}, (2001).} \label{f:1} 
\end{figure}

\section{Structures in the Local Interstellar Medium}

The D/H ratio is unlikely to be constant throughout the Galaxy. Prime
candidates for different D/H ratios are those locations where the gas has been
confined for a long time and the gas composition has been altered by stellar
mass loss of astrated material with limited mixing with the gas in the rest of
the Galaxy. We do not know {\em a priori} what these structures are, but as a
start we should measure the D/H ratio in ISM gas located in identifiable
structures in our local region of the Galaxy. 

The Galactic halo extends for many kiloparsecs (kpc) above and below the
Galactic plane and likely consists primarily of hot gas with low metal
abundances. A prime goal of the Far Ultraviolet Spectrograph Explorer (FUSE)
mission is to measure D/H in sightlines through the halo, but there are no
results available to report. The thin disk of the Galaxy, in which most giant
molecular clouds are located and star formation occurs, has a vertical scale
height of 325~pc and a radial scale height $\sim 4000$~pc. The D/H ratios
measured toward OB stars in the thin disk by the {\em Copernicus} satellite and
the IMAPS instrument will be discussed below. 

The Sun is located inside a region of very low density called the Local Cavity.
\inlinecite{Sfeiretal99} have modelled the contours of the Na~I absorption that
likely delineate the outer edge of hot low density gas ($\log T =$ 6.0--6.1)
called the Local Bubble (LB), which extends outward by 100--200~pc from the
Sun. It is likely that the LB fills most or all of the Local Cavity, but this
is not yet demonstrated. The LB was likely formed by the winds and supernovae
explosions of stars in the Scorpius-Centaurus Association as the 26 km s$^{-1}$
flow vector is from the center of the Association. The age of the LB is a few
million yr and the gas within it is likely well mixed and could be D-poor and
metal-rich given its origin. 

Within the LB are a number of small clouds consisting of warm, partially
ionized gas (see Fig. 2). The Sun is located within but close to the edge of
the Local Interstellar Cloud (LIC). First identified from its kinematics by
\inlinecite{LallementBertin92}, the LIC was modelled by
\inlinecite{RedfieldLinsky00} as roughly spherical with dimensions of 5--8 pc,
$T\approx 7,000$~K, and $n_{total}\approx 0.2$ cm$^{-3}$. Within the LIC, D/H
and the depletions of Mg and Fe appear to be constant. Located near the LIC are
at least nine other warm clouds with similar temperatures but a wide range of
metal depletions, indicating that grains in some clouds have been evaporated by
shocks. The ionization fractions of H and He in the LIC are consistent with
steady-state equilibrium for which the photoionization is from nearby stars
(primarily $\epsilon$~CMa), the UV background, and an assumed UV radiation
field formed at the boundary between warm clouds and the hot surrounding gas
(cf. model 17 in \opencite{Slavinfrisch02} and \opencite{Woodetal02b}). 

\begin{figure}
\centerline{\psfig{file=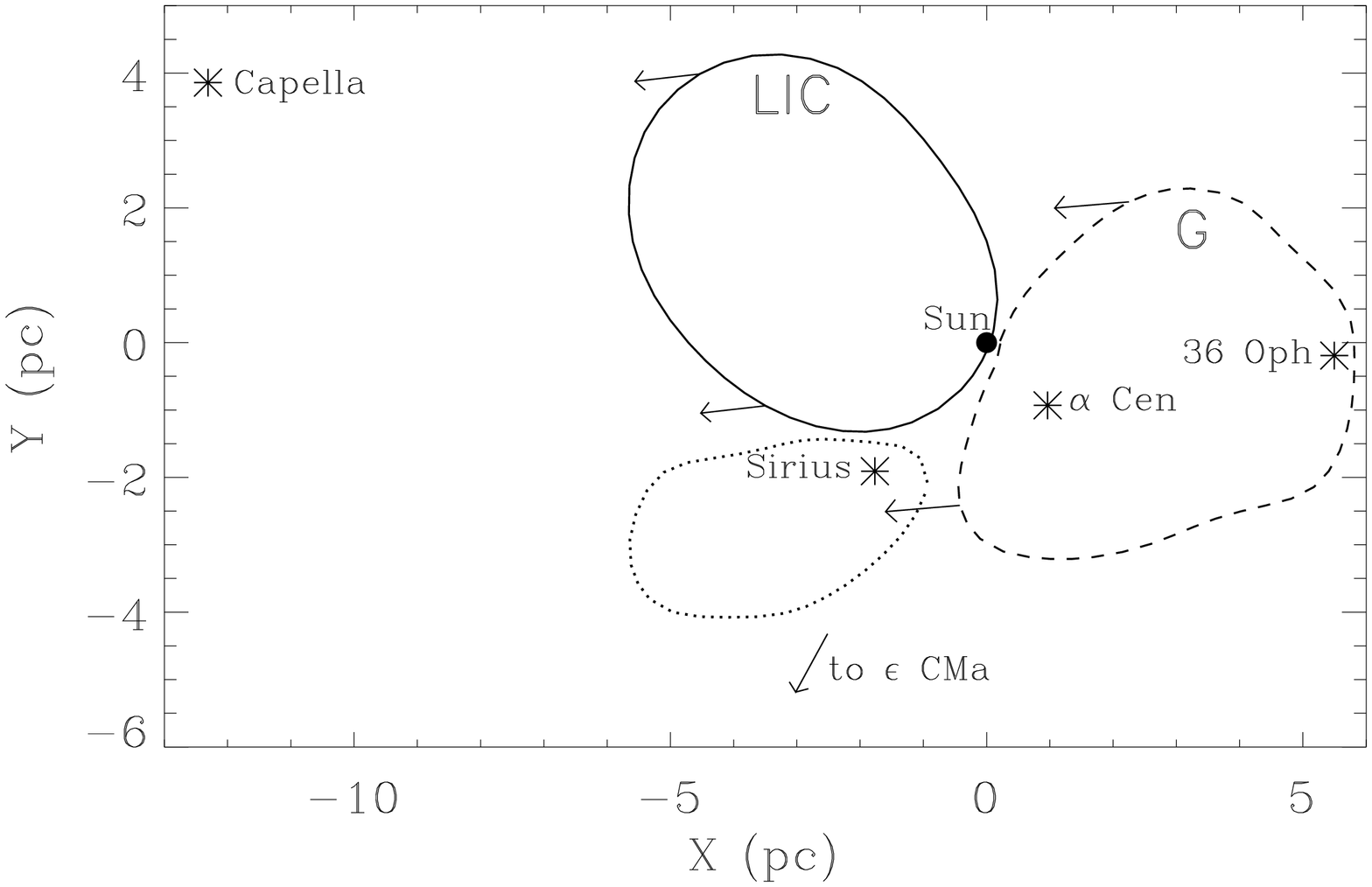,width=0.95\textwidth,clip=,angle=90}
\vspace{0.1cm}}
\caption{A schematic view of the Local Interstellar Cloud (LIC) and two other 
clouds as viewed from the North Galactic Pole. The Sun is located just inside 
the LIC toward the G cloud. Arrows designate the ISM flow direction. The star 
$\epsilon$~CMa is a major source of the photoionizing radiation. From 
Wood {\it et al.}, (2002b).} \label{f:2}
\end{figure}
\vspace{-11pt}

\section{FUSE Measurements of D/H Along the Lines of Sight to Nearby Hot White
Dwarf Stars} 

The FUSE spacecraft obtains spectra of stars and extragalactic sources in the
far--UV (910--1180~\AA) with about 20 km s$^{-1}$ resolution. For a description
of the satellite and its capabilities see \inlinecite{Moosetal00} and
\inlinecite{Sahnowetal00}. A major goal of the FUSE observing program is to
measure D/H in local and more distant interstellar gas. The first results of
this program are published in a series of eight papers appearing in the May
2002 issue of ApJ Supplements. \inlinecite{Moosetal02} summarize the results
obtained from analyses of the lines of sight to five white dwarfs (HZ~43,
G191-B2B, WD~0621-376, WD~1634-573, and WD~2211-495) located at distances of
37--69~pc within the LB and to two subdwarfs (Feige~110 and
BD~+28$^{\circ}$4211) located at distances of 104--179~pc outside of the LB.
White dwarfs are useful targets because they have a bright continuum with few
stellar absorption lines, relatively simple lines of sight, and no stellar 
winds.

The basic approach taken in analyzing these spectra is to fit Voigt (convolved
Doppler and damping) profiles to the interstellar absorption seen in the FUSE
and STIS spectra to determine the number of absorption components and the total
column densities for H, D, and important metals. One complication is that the
FUSE spectra have insufficient spectral resolution to determine N(H~I) from the
shapes of the higher Lyman lines, so N(H~I) is better determined from the
Lyman-$\alpha$ profile or EUVE measurements of the Lyman continuum absorption.
Uncertainties in the FUSE line spread function and velocity scale also
complicate the analysis. The possible presence along the line of sight of hot
hydrogen absorbers with low column densities increases the uncertainty in
N(H~I) and thus D/H. 

\begin{table}
\caption{FUSE results for D/H from D/O and O/H\vspace{-0.1cm}}
\begin{tabular*}{12.5cm}{lcc}
\hline \vspace{-0.08cm}
Number Ratio & 5 sightlines inside LB        & All 7 sightlines\\ \hline
D~I/O~I (FUSE LISM) 
             & $(3.76\pm 0.20)\times 10^{-2}$& $(3.99\pm 0.19)\times 10^{-2}$\\
O/H (Sun)    & $(4.90\pm 0.56)\times 10^{-4}$& $(4.90\pm 0.56)\times 10^{-4}$\\
D/H (ppm)    & {\bf 18.4}$\pm${\bf 2.3}      & {\bf 19.5}$\pm${\bf 2.5} \\
D/H (O$_{gas}$/O$_{tot}=0.80$)
             & {\bf 14.7}$\pm${\bf 1.8}      & {\bf 15.6}$\pm${\bf 2.0} \\
             &                               &                 \\
O~I/H~I (ISM)&$(3.43\pm 0.15)\times 10^{-4}$ & $(3.43\pm 0.15)\times 10^{-4}$\\
D/H (ppm)    & {\bf 12.9}$\pm${\bf 0.8}      & {\bf 13.7}$\pm${\bf 0.8} \\
             &                               &                  \\
O~I/H~I (FUSE LISM) 
             &$(3.94\pm 0.35)\times 10^{-4}$ & $(3.03\pm 0.21)\times 10^{-4}$\\ 
D/H (ppm)    & {\bf 14.8}$\pm${\bf 1.5}      & {\bf 12.1}$\pm${\bf 1.0}\\ \hline
\end{tabular*}
\vspace{-0.6cm}
\vspace{-22pt}
\end{table}

An example of this work is the line of sight to G191-B2B analyzed by
\inlinecite{Lemoineetal02}. The line of sight to this hot DA white dwarf ($\log
T_{\rm eff} = 54,000$~K, $\log g = 7.4$) has 3 ISM velocity components: 19.6
km/s (LIC), 11.5 km/s, and 7.4 km/s. The inclusion of uncertainties in N(H~I)
from the possible presence of hot H absorbers and uncertainties in the stellar
Lyman line shapes against which the interstellar absorption is measured leads
to log N(H~I) = $18.18 \pm 0.18$ (2 $\sigma$) and (D/H)$_{tot}$ = {\bf
16.6}$^{\bf +9}_{\bf -6}$ ppm. 

For all seven lines of sight typical uncertainties in N(D~I) are $\pm$10\%
(1$\sigma$), but the values of N(H~I) obtained from EUVE, GHRS, STIS, or IUE
spectra are typically uncertain by $\pm$17\% (1$\sigma$). The weighted mean D/H
= {\bf 15.2} $\pm$ {\bf 0.8} ppm and the range in D/H values is {\bf 14--21}
ppm. The line of sight with the highest D/H = {\bf 21.4}$\pm${\bf 4.1} ppm is
Feige 110, which is located ouside of the LB. D/H for the other six lines of
sight cluster closely about {\bf 15} ppm. 

An alternative and perhaps more accurate way of determing D/H is from
measurements of D/O and O/H. Oxygen is a good proxy for hydrogen as the
ionization potentials for O~I and H~I are nearly the same and their ionization
equilibria are closely tied by charge exchange reactions. The presence of many
optically thin O~I lines in the FUSE spectrum leads to typical uncertainties in
N(O~I) of $\pm$10\%. Typical uncertainties in D/O are $\pm$15\%, and for the
five white dwarfs inside the LB the weighted mean value is D/O = 0.0376 $\pm$
0.0020 ($\pm$5\%). The usually cited value is O~I/H~I = $(3.43\pm 0.15)\times
10^{-4}$ \cite{Meyer02} for the ISM at distances of 200--1000~pc. I take the
new solar value of O/H = $(4.90\pm 0.56)\times 10^{-4}$ from
\inlinecite{Allende2001}, who use their three-dimensional time-dependent
hydrodynamical model solar atmospheres to analyze the [O~I] 6300~\AA\ line.
Table I summarizes the D/H ratios derived using the measured D/O ratios and
either the O/H ratio for the Sun (assuming 0\% or 20\% depletion of oxygen on
to grains in the ISM), the Meyer (2002) value for the ISM gas, or the FUSE
values for the LISM gas. For nearly all of these cases, the inferred D/H value
is consistent with the directly measured value of D/H = {\bf 15.2} $\pm$ {\bf
0.8} ppm. 

\section{What have We Learned about D/H in the Galaxy?}

\begin{table}
\caption{D/H in the Local Bubble measured by different techniques
\vspace{-0.1cm}}
\begin{tabular*}{12.5cm}{lcc}
\hline \vspace{-0.08cm}
Measurement Procedure      & D/H (ppm)            & Reference \\ \hline
HST (solar-like stars within 100~pc) & {\bf 15.0}$\pm${\bf 1.0} &
\inlinecite{Linsky98}\\
FUSE (5 sightlines inside LB) &  {\bf 15.2}$\pm${\bf 0.8}  & 
\inlinecite{Moosetal02}\\
FUSE (D/O and O/H in LISM) & {\bf 14.8}$\pm${\bf 1.5} & 
\inlinecite{Moosetal02}\\
FUSE (D/O, solar O/H)      & {\bf 18.4}$\pm${\bf 2.3} & Table~I\\
FUSE (D/O, solar O/Hx0.80)& {\bf 14.7}$\pm${\bf 1.8} & Table~I\\
\hline
\end{tabular*}
\vspace{-0.6cm}
\end{table}

\begin{table}
\caption{D/H measurements beyond the Local Bubble\vspace{-0.1cm}}
\begin{tabular*}{12.5cm}{lcccc}
\hline \vspace{-0.08cm}
Star     &  D (pc)    & D/H (ppm)  &   Instrument   & Reference  \\ \hline
BD+28$^{\circ}$4211 & 104  & {\bf 13.9}$\pm${\bf 1.0} & FUSE & 
\inlinecite{Sonnebornetal02}\\
$\theta$ Car & 135 & {\bf 5.0}$\pm${\bf 1.6} & Copernicus & 
\inlinecite{Allenetal92}\\
Feige 110 & 180 & {\bf 21.4}$\pm${\bf 4.1} & FUSE & 
\inlinecite{Friedmanetal02}\\
$\gamma$ Cas & 188 & {\bf 13}$\pm${\bf 2.5} & Copernicus & 
\inlinecite{Ferletetal80}\\
$\lambda$ Sco & 216 & {\bf 7.6}$\pm${\bf 2.5} & Copernicus & 
\inlinecite{York83}\\
$\gamma^2$ Vel & 258 & {\bf 21.8}$\pm${\bf 2.0} & IMAPS & 
\inlinecite{Sonnebornetal00}\\
$\zeta$ Pup & 430 & {\bf 14.2}$\pm${\bf 1.5} & IMAPS & 
\inlinecite{Sonnebornetal00}\\
$\delta$ Ori A & 500 & {\bf 7.4}$\pm${\bf 1.0} & IMAPS & 
\inlinecite{Jenkinsetal99}\\
4 QSOs    &    & {\bf 30}$\pm${\bf 4} & Keck & \inlinecite{Omearaetal01}\\
\hline
\end{tabular*}
\vspace{-0.6cm}
\end{table}

\begin{itemize}
\item[$\bullet$] Within the Local Bubble (out to 100~pc or more from the Sun),
D/H probably has a single value (i.e., the local ISM is well mixed). The
measurements of D/H summarized in Table~II lead me to conclude that the best
value for D/H in the Local Bubble is (D/H)$_{LB}$ = {\bf 15}$\pm$ {\bf 1} ppm. 

\item[$\bullet$] Table~III and Figure 3 summarize the D/H measurements of gas
beyond the LB, including measurements toward two hot subdwarfs by FUSE, to
three O stars by Copernicus, to three O stars by the IMAPS experiment, and the
mean of four quasar sightlines studied with the Keck telescope. These results
show a wide range of D/H = {\bf 5--22} ppm in the Galactic disk. 

\item[$\bullet$] If we adopt the most recent quasar sightline value of D/H =
{\bf 30$\pm$ 4} ppm \cite{Omearaetal01} as an approximate value for
(D/H)$_{prim}$, then the deuterium astration in the Local Bubble,
(D/H)$_{prim}$/(D/H)$_{LB}$ = ($30\pm 0.4$)/($15\pm 1$) = $2.0\pm 0.4$. The
range of deuterium astration in the Galactic disk from the data in Table~III is
then 1.35--6.0. 

\item[$\bullet$] Theoretical estimates of deuterium astration over the lifetime
of our Galaxy are $\leq 3$ \cite{Tosietal98} and appear to be inconsistent with
the wide range of observed astration values. However, the models make a number
of assumptions that may not be valid. For example, the young Galaxy has
primordial D/H and no metals, and the infalling gas from the halo has
primordial or near-primordial abundances. Each ring of the Galaxy (several kpc
wide) is assumed to be well mixed, and the gas is not mixed with gas in other
rings. If the D/H results beyond the Local Bubble are valid, the Galactic
chemical evolution models are \underline{overly simplified}. The next
generation of Galactic chemical evolution models must include episodic star
formation (star bursts) with rapid mass loss and supernovae events and more
realistic mixing scenarios. 

\item[$\bullet$] All Galactic chemical evolution models predict that D/H and
metal abundance should be anti-correlated, but the initial results from FUSE do
not show this. Rather, there appears to be a weak positive correlation between
D/H and O/H. Analysis of D/H and O/H for more lines of sight is needed. 

\item[$\bullet$] (D/H)$_{prim} \approx$ (D/H)$_{QSO} =$ {\bf 30}$\pm$ {\bf 4}
ppm is consistent with the primordial abundance of He and $^7$Li according to
present models of Big Bang nucleosynthesis (e.g., \opencite{Burlesetal01}). The
ratio of baryons to photons is $\eta = (5.5\pm 0.5)\times 10^{-10}$, and the
ratio of the baryon density to the closure density, $\Omega_B = 0.041\pm
0.009$. Big bang nucleosynthesis theory looks basically right. 

\end{itemize} 

\begin{figure}
\centerline{\psfig{file=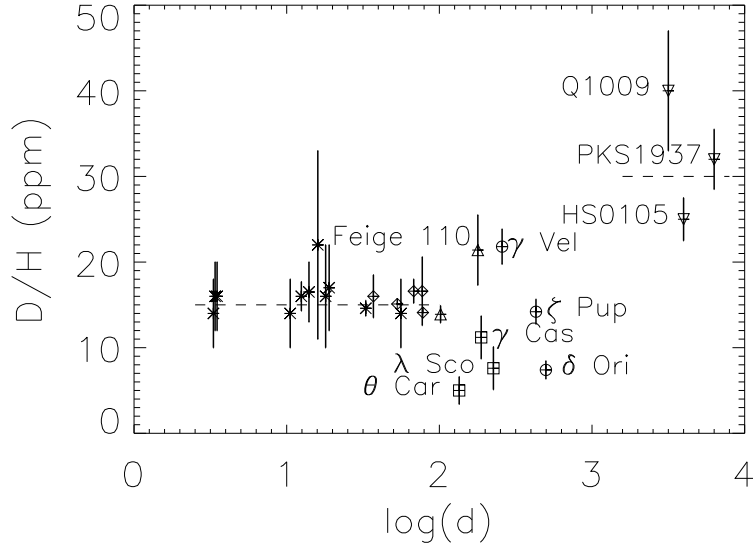,width=0.95\textwidth,clip=}
\vspace{0.1cm}}
\caption{A summary of D/H measurements obtained with HST (asterisks), FUSE
inside the LB (diamonds), FUSE outside of the LB (triangles), Copernicus
(squares), IMAPS (circles), and Keck (upside down triangles) with distance d in
units of parsecs. The three quasar lines of sight studied with Keck are not
plotted at their correct distances. The dashed lines refer to the mean D/H
values inside the Local Bubble and for the quasar lines of sight.} \label{f:3} 
\end{figure}

\begin{acknowledgements}
This work is supported by NASA grant S-56500-D to the University of Colorado 
and NIST and contract NAS5-32985 to The Johns Hopkins University for the 
NASA--CNES--CSA FUSE project. I would also like to thank Greg Herczeg, Seth 
Redfield, and Brian Wood for discussions, support, and figures.

\end{acknowledgements}
\vspace{-11pt}


\end{article}
\end{document}